# Retransmission Steganography Applied

Wojciech Mazurczyk, Miłosz Smolarczyk, Krzysztof Szczypiorski
Institute of Telecommunications
Warsaw University of Technology
Warsaw, Poland

e-mail: wmazurczyk@tele.pw.edu.pl, milosz.smolarczyk@gmail.com, ksz@tele.pw.edu.pl

*Abstract* — **This paper presents experimental results of the implementation of network steganography method called RSTEG (Retransmission Steganography). The main idea of RSTEG is to not acknowledge a successfully received packet to intentionally invoke retransmission. The retransmitted packet carries a steganogram instead of user data in the payload field. RSTEG can be applied to many network protocols that utilize retransmissions. We present experimental results for RSTEG applied to TCP (Transmission Control Protocol) as TCP is the most popular network protocol which ensures reliable data transfer. The main aim of the performed experiments was to estimate RSTEG steganographic bandwidth and detectability by observing its influence on the network retransmission level.**

## I. INTRODUCTION

Steganography includes information hiding techniques to deliver secret data (steganograms) from sender to the receiver in such a way that no one else would be aware of the message existence and its exchange. To achieve these goals it needs a carrier in which steganograms will be embedded. Typical steganographic methods utilized as a carrier digital media like images, audio or video. Recent trend in information hiding is network steganography. It creates hidden channels by modifying network protocols. Thus in this case network protocol is a carrier of the secret data. Protocol-carrier modification includes modification of PDU (Protocol Data Unit), modification of PDU time relations or both (hybrid solution).

RSTEG (Retransmission Steganography) is a hybrid network steganography method, which is intended for a broad class of protocols that utilise retransmission mechanisms. The main innovation of RSTEG is to not acknowledge a successfully received packet in order to intentionally invoke retransmission. The retransmitted packet of user data then carries a steganogram in the payload field.

In 2009 we introduced RSTEG [4] and we presented simulation results which main aim was to measure and compare steganographic bandwidth of the proposed method for different TCP retransmission mechanisms as well as to determine the influence of RSTEG on the network retransmissions level. This paper is an extension of this work as we focus on practical RSTEG applications.

RSTEG can be applied to many network protocols utilizing retransmissions. We decided to perform experiments for TCP (Transmission Control Protocol) [5] as a vast amount of Internet traffic (about 80-90%) is based on this protocol.

Previous experiments were carried out in network simulators: *ns2* and *ns3* [10]. They have shown that RSTEG is effective and hard to detect. Results from these two simulation environments were different. However, it is not surprising as they both have different implementations of complex network protocols such as TCP. To prove RSTEG effectiveness and provide reliable results we decided to implement RSTEG on a Linux kernel and measure network traffic for the real TCP/IP stack. Main idea of RSTEG and then achieved experimental results are presented and described in next sections.

## II. GENERAL IDEA OF RSTEG

In a simplified situation, a typical protocol that uses a retransmission mechanism based on timeouts obligates the receiver to acknowledge each received packet. When the packet is not successfully received, no acknowledgment is sent. After the timeout expires and the sender has not received the acknowledgement the packet is retransmitted (Fig. 1).

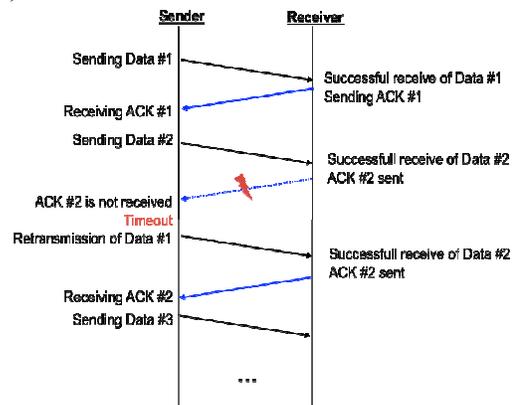

**Figure 1.** Generic retransmission mechanism based on timeouts

As mentioned in Section I, RSTEG uses a retransmission mechanism to exchange steganograms. Both the sender and the receiver are aware of the steganographic procedure. They reliably exchange packets during their connection; that is, e.g. they transfer a file. At some point during the connection, when the sender wants to send a steganogram, the receiver after successfully receiving a packet intentionally does not issue an acknowledgment message. In a normal situation, the sender is obligated to retransmit the lost packet when the timeframe, within which packet acknowledgement should have been received, expires. In the context of RSTEG, a sender replaces original payload with a steganogram instead of sending the same packet again. When the retransmitted packet reaches the receiver, he/she can then extract hidden information (Fig. 2).

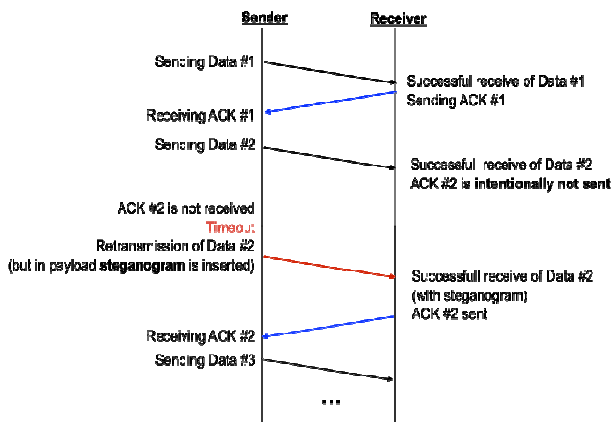

**Figure 2.** RSTEG main idea

The intentional retransmissions due to RSTEG should be kept at a reasonable level to avoid detection. To achieve this goal, it is necessary to determine the average number of natural retransmissions in TCP-based Internet traffic as well as to know how intentional retransmissions affect the network retransmission rate. Usually network retransmissions are caused by network overload, excessive delays or reordering of packets [6], and their number is estimated to account for up to 7% of all Internet traffic [6], [3], [1].

It must be also noted that based on results presented in [9] up to 0.09 % (1 in 1100) of TCP segments may be corrupted due to network delivery. As a result, an imperfect copy of a segment may be sent to the receiver. After reception of the invalid segment, verification is performed based on the value in the *TCP Checksum* field, and the need to retransmit is signalled to the sender. Thus, in this scenario, the original segment and the retransmitted one will differ from each other. Occurrences of this effect in IP networks mask the use of RSTEG.

It is worth noting that even for the low rates of intentional retransmission (0.09%) that are required to mask RSTEG, if we assume that the TCP segments are generated at a rate of 200 segments/s, with the connection lasting 5 minutes and the segment's payload size being 1000 bytes, then this results in $S_B$ = 180 Bps, which is a rather high bandwidth, considering the other steganographic methods.

RSTEG can be applied to all retransmission mechanisms used in TCP namely RTO (Retransmission Timeout) FR/R (Fast Retransmit/Recovery) or SACK (Selective ACK). It requires modification to both the sender and to the receiver. The sender should control the insertion procedure and decide when a receiver should invoke a retransmission. The sender is also responsible to keep the number of retransmissions at a non-suspicious level. The receiver's role is to detect when the sender indicates that the intentional retransmission should be triggered. Then, when the retransmitted segment arrives, the receiver should be able to extract the steganogram.

The sender must be able to mark segments selected for hidden communication (that is, retransmission request segments) so the receiver would know which segments retransmissions should be invoked and which segments will contain steganograms. However, marked TCP segment should not differ from those sent during a connection. The following procedure for marking sender segments is proposed. Let us assume that the sender and receiver share a secret Steg-Key (*SK*). For each fragment chosen for steganographic communication, the following hash function (*H*) is used to calculate the Identifying Sequence (*IS*):
Note that *Sequence Number* and *TCP Checksum* denote

$$IS = H(SK \parallel Sequence\ Number \parallel TCP\ Checksum \parallel CB)$$

values from the chosen TCP header fields in segments, ‖ is the bits concatenation function, and *CB* is a control bit that allows the receiver to distinguish a retransmission request segment from a segment with a steganogram. For every TCP segment used for hidden communication, the resulting *IS* will have different value due to the variety of values in the *Sequence Number* and *TCP Checksum* header fields. All *IS* bits (or only selected ones) are distributed by the sender across a segment's payload field in a predefined manner. The receiver must analyse each incoming segment; based on *SK* and values from the TCP header, the receiver calculates two values of *IS*, namely, one with *CB* = 1 and one with *CB* = 0. Then the receiver checks if and which *IS* is present inside the received segment.

Let us present how RSTEG may be applied to the RTO retransmission mechanism (Fig. 2):

- The sender marks a segment selected for hidden communication by distributing the *IS* across its payload.
- After successful segment delivery, the receiver does not issue an ACK message.
- When the RTO timer expires, the sender sends a steganogram inside the retransmitted segment's payload.
- The receiver extracts the steganogram and sends the appropriate acknowledgement.

Problems may arise when the segment that informs the receiver of a necessity to invoke an intentional retransmission (which contains user data together with the *IS*) is lost due to network conditions. In that case, a normal retransmission is triggered, and the receiver is not aware that the segment with hidden data will be sent. However, in this case, the sender believes that retransmission was invoked intentionally by the receiver, and so he/she issues the segment with steganogram and the *IS*. In this scenario, user data will be lost, and the cover connection may be disturbed.

In order to address the situation in which the receiver reads a segment with an unexpected steganogram, the receiver should not acknowledge reception of this segment until he/she receives the segment with user data. When the ACK is not sent to the sender, another retransmission is invoked. The sender is aware of the data delivery failure, but he/she does not know which segment to retransmit, so he/she first issues a segment with user data. If delivery confirmation is still missing, then the segment with steganogram is sent. The situation continues until the sender receives the correct ACK (Fig. 3).

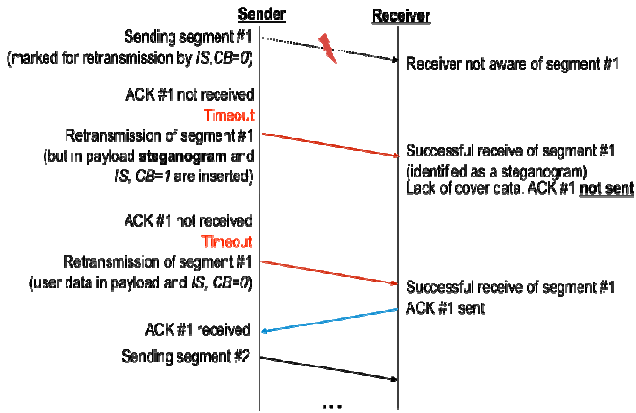

**Figure 3** RTO-based RSTEG segment recovery example

For example, consider the scenario in which we invoke 0.5% intentional retransmissions. If 5% is lost, it means that the above-described mechanism will take place only for 0.025% of steganogram segments, and thus it will be used rarely.

### III. RSTEG IMPLEMENTATION

Experimental RSTEG implementation has been done on Linux 2.6.27.7-9 kernel. It allows to measure steganographic bandwidth, retransmission difference and other values presented in Section V. Some important modifications to the TCP/IP stack are described below.

#### A. Sending procedure modifications

Sending phase was modified to queue steganograms delivered by application layer and wait for cover data (which also come from the application). When cover data comes to Linux kernel, *IS (Identifying Sequence)* is inserted to mark segment for retransmission. In this experimental version predefined data is used as *IS* to recognize it easily in network traffic dumps.

The Linux TCP/IP stack has some data transfer and kernel operations' optimizations e.g. data collation, putting data in many blocks in memory [7]. These optimizations cannot be used with RSTEG. The first one could cause joining segments containing secret and cover data. Joining only cover or secret data in one segment does not affect steganographic transmission, but in this experiment we turned off this mechanism to simplify procedure. The second optimization mechanism is used with network cards that support scatter-gather operations, in other cases all user data is linearized by kernel. Changing data, which is split in many memory blocks, require complex operations, so we decided to always linearize user data.

#### B. Receiving procedure modifications

Receiver's task is to recognize segments containing *IS*. The sequence is computed for each incoming segment and compared with extracted one. If segment is recognized as RSTEG retransmission request, then no ACK is sent, otherwise data is acknowledged and delivered to an application.

If steganogram arrives and receiver detects lack of retransmission request for this segment then no ACK is sent and recovery procedure is applied (see Section II).

Among all kernel optimization mechanisms only adjacent segment collapsing is unable to work with RSTEG. Adjacent segment collapsing joins data from many segments into one block before delivering to application. RSTEG requires delivering segments separately because steganograms are recognized also in application layer.

#### C. Retransmission procedure modifications

Retransmission procedure is the most important phase of steganographic communication. For each segment marked for retransmission by RSTEG, retransmission counter is created. If retransmission is triggered and counter is zero or even then the payload is replaced by steganogram, otherwise segment is retransmitted without change (recovery procedure).

After data replacement it is necessary to update also the checksum unless network card supports *TCP Checksum Offload*, which processes checksum calculation on the network card.

### IV. EXPERIMENT METHODOLOGY

The network topology (Fig. 4) was designed to fit Internet traffic retransmission statistics (see Section II). The SRC node transmits TCP traffic and UDP background traffic to DEST node through 1 Mb/s bottleneck, which causes natural retransmissions.

Traffic parameters which are presented in table below were matched to achieve ~3-4% of natural retransmissions with zero size buffers on routers R1 and R2.

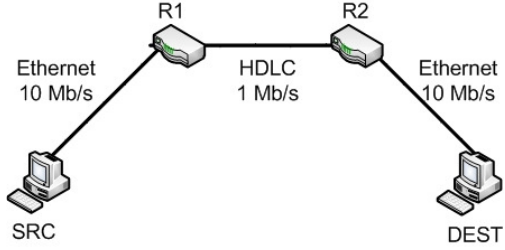

**Figure 4. Experimental network topology**

TABLE I. EXPERIMENTAL TRAFFIC PARAMETERS

| Parameter name | Value |
|---|---|
| TCP throughput | 125 kb/s |
| UDP throughput | 1 160 kb/s |
| Transmission time | 600 s |
| Measured time | 540 s |
| Measure start delay | 60 s |
| Payload size | 1200 B |

Network traffic was measured for 9 minutes, starting after 1 minute from the beginning of transmission. The RSTEG intentional retransmission probability ($IR_P$) was changed from 0 to 5% with intermediary steps at 0%, 1%, 2%, 3%, 4% and 5%.

In the above simulation scenario, five parameters were measured:

- **Steganographic Bandwidth ($S_B$)** - the amount of the steganogram transmitted using RSTEG during one second [B/s]. Parameter may be expressed as

$$S_B = \frac{N_S \cdot S_S}{T} \quad [Bps]$$

where:
$N_S$ – the number of segments used for hidden communication
$S_S$ – the size of segment payload
$T$ – the duration of the connection

$S_B$ was measured by receiver`s application, which counted number of the steganograms and their size.

- **Retransmissions Difference ($R_D$)** - the difference between retransmissions in a network after applying RSTEG and in a network before applying RSTEG. This parameter can be used to estimate the influence that RSTEG has on the TCP retransmissions rate. Thus, it can illustrate how to choose the correct intentional retransmission probability to limit the risk of detection.

$R_D$ was measured with Wireshark sniffer (www.wireshark.org) in pcap traffic dumps by counting segments suspected to be retransmitted.

- **Steganographic Retransmissions Ratio ($S_R$)** - the amount of steganographic retransmissions in all retransmissions.

- **TCP Throughput ($T_T$)** - the effective TCP throughput, measured on the DEST node.

- **Effective $IR_P$** - $IR_P$ measured on the DEST node.

V. EXPERIMENTAL RESULTS

The results achieved in the experiment are presented in Fig. 5 and 6.

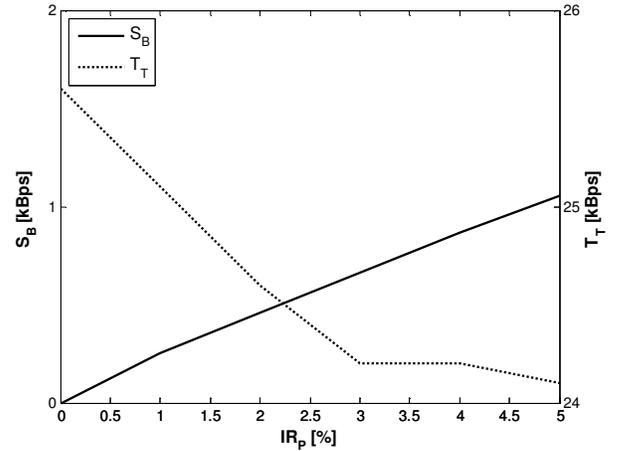

**Figure 5. Steganographic Bandwidth ($S_B$) and TCP Throughput ($T_T$)**

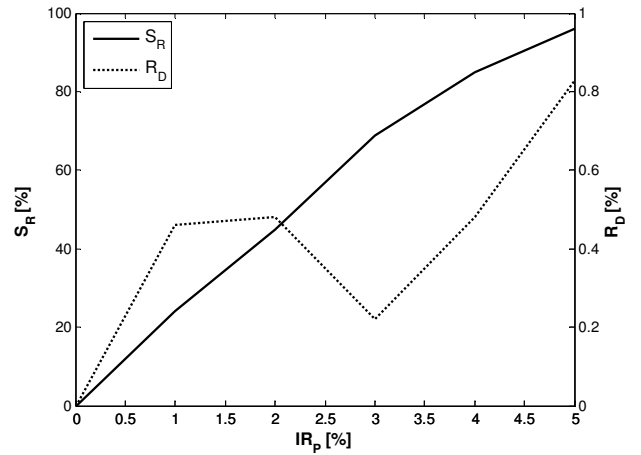

**Figure 6. Steganographic Retransmissions Ratio ($S_R$) Retransmissions Difference ($R_D$)**

Table 2 summarizes achieved experimental results.

TABLE II. SUMMARIZED EXPERIMENTAL RESULTS

| $IR_P$ | $S_B$ | | $R_D$ | | $S_R$ | | TCP throughput | | Effective $IR_P$ | |
|---|---|---|---|---|---|---|---|---|---|---|
| | [B/s] | σ | [%] | σ | [%] | σ | [kB/s] | σ | [%] | σ |
| 0% | 0 | 0 | 0 | 0 | 0 | 0 | 25.6 | 0.2 | 0 | 0 |
| 1% | 252 | 18 | 0.46 | 0.23 | 24 | 2.9 | 25.1 | 0.2 | 1.01 | 0.08 |
| 2% | 461 | 37 | 0.48 | 0.23 | 45 | 4.3 | 24.6 | 0.1 | 1.87 | 0.15 |
| 3% | 665 | 37 | 0.22 | 0.18 | 69 | 3.4 | 24.2 | 0.2 | 2.74 | 0.16 |
| 4% | 867 | 36 | 0.48 | 0.16 | 85 | 5.1 | 24.2 | 0.1 | 3.58 | 0.15 |
| 5% | 1056 | 58 | 0.83 | 0.2 | 96 | 2.3 | 24.1 | 0.1 | 4.39 | 0.24 |

The results show that bandwidth of the steganographic channel is increasing together with intentional retransmissions, however increase of $R_D$ is slower (Fig. 5). In real world TCP/IP stack implementation the congestion avoidance algorithms ([7], [8]) are reducing congestion window, which is natural response to retransmissions. That effect causes throughput reduction and smaller retransmissions difference than was intentionally triggered in [1%; 2%] $IR_P$ range. When $IR_P$ reaches 3% its value is around number of retransmissions for normal network conditions (RSTEG-free) and $R_D$ is non-zero because of random choice of segments marked for retransmission. Next, the $IR_P$ reaches up to 5%, which causes more retransmissions in the network again, but still fewer than number of the intentional retransmissions because of congestion avoidance algorithms.

Fig. 6 shows that in network congestion state intentional retransmissions are reducing the number of natural ones (but of course total number of retransmissions is still higher). The optimal intentional retransmissions level is 5% when almost all retransmitted segments are used for steganographic communication.

Keeping total retransmissions number on reasonable level is necessary to avoid detection. The experimental results show that we can achieve decent steganographic bandwidth while maintaining non-suspicious level of retransmissions.

## VI. CONCLUSIONS

RSTEG is a hybrid network steganography method which to transmit steganograms utilizes intentionally invoked retransmission.

In this paper we presented how RSTEG can be integrated into Linux kernel and we showed the first experimental results made on real TCP/IP stack.

The results proved that RSTEG is an effective steganographic method which offers relatively high steganographic bandwidth when compared with other network steganography methods [4]. The steganographic bandwidth depends on RSTEG intentional retransmissions level.

The results have shown that RSTEG intentional retransmissions level affects number of usually retransmitted segments during connection. Keeping reasonable total retransmissions level is very important because of RSTEG detection potential methods. RSTEG steganalysis method may be implemented with a passive warden [2] (or some other network node responsible for steganography usage detection). Passive warden must be able to monitor all the TCP traffic and for each TCP connection it must store sent segments for the given period of time, which depends on the retransmission timer i.e. passive warden must store the segment until it is acknowledged by the receiver so the retransmission is not possible any more. When there is a retransmission issued, passive warden compares originally sent segment with retransmitted one and if the payload differs RSTEG is detected and the segment is dropped. However, it should be noted that there may be serious performance issues involved if passive warden monitors all the TCP connections and must store a large number of the segments. [4]. Moreover the RSTEG causes less increase of total retransmissions level than $IR_P$ level which makes it harder to detect. Of course it is still needed to keep intentional retransmissions level on reasonable level.